\documentstyle[prc,aps,floats,psfig,twocolumn]{revtex}

\draft

\begin{document}

\title{
Scaling laws and transient times in $^3$He induced
nuclear fission
}
\author{
 Th.~Rubehn,\thanks{Electronic address: TRubehn@lbl.gov}
 K.X.~Jing, 
 L.G.~Moretto, 
 L.~Phair, 
 K. Tso, 
 and G.J.~Wozniak }
\address{
Nuclear Science Division,
Lawrence Berkeley National Laboratory,
University of California, 
Berkeley, California 94720
}
\date{\today}
\maketitle

\begin{abstract}
Fission excitation functions of compound nuclei in a mass
region where shell effects are expected to be very strong
are shown to scale exactly according to the transition state 
prediction once these shell effects are accounted for. 
The fact that no deviations from the transition state 
method have been observed within the experimentally
investigated excitation energy regime 
allows one to assign an upper limit for the 
transient time of 10$^{-20}$ seconds. 
\end{abstract}


\section{Introduction}
More than half a century after its discovery \cite{Fis},
the study of fission is still of general interest. 
While the availability of relativistic heavy ions
has enabled the study of
several aspects of the fission process
in the high energy region 
\cite{Sch94,Ber94,Pol94,Cle95,Ign95,Rub95,Arm95},
it has been shown recently that a new approach \cite{Mor95} to
investigate excitation functions of low energy, light 
particle induced fission allows for the model independent
extraction of fundamental quantities of the fission process,
like fission barriers, shell effects, and the much
discussed fission delay time
(see e.g. Refs.~\cite{Mor95,Hil92,Pau94}).

From early studies it is well known that the
fission excitation functions vary dramatically
from nucleus to nucleus over the periodic table
\cite{Rai67,Mor72,Kho66}:
Some of the differences can be understood in terms of a changing
liquid-drop fission barrier with the fissility parameter, 
others are due to to strong shell
effects which occur e.g. in the neighborhood of the double magic 
numbers $Z$=82 and $N$=126. 
Further effects may be associated with pairing and the angular 
momentum dependence of the fission barrier \cite{Van73,Wag91}.

Fission rates have been calculated most often on the basis of
the transition state method introduced by
Wigner \cite{Wig38}, and applied to fission by
Bohr and Wheeler \cite{Boh39}.
The success of this method has prompted attempts to justify
its validity in a more fundamental way, and to identify
regimes in which deviations might be expected.
Recent publications claim
the failure of the transition state rates to account for the 
measured amounts of prescission neutrons or $\gamma$-rays in
relatively heavy fissioning systems 
\cite{Hil92,Pau94,Tho93}. 
This alleged failure has been attributed to the transient time
necessary for the so-called slow fission mode to attain its
stationary decay rate 
\cite{Gra83a,Gra83b,Wei84,Gra86,Lu86,Lu90,Cha92,Fro93,Siw95}.
The larger this fission delay time, the more favorably
neutron decay competes with the fission process. 
This leads to an effective fission probability smaller
than predicted by the Bohr - Wheeler formula.
The experimental methods of these studies, however, suffer from two
difficulties: First they require a possibly large correction
for post-saddle, but pre-scission emission; second, they
are indirect methods since they do not directly determine
the fission probability. 
The measured prescission particles can be emitted either
before the system reaches the saddle point, or during
the descent from saddle to scission. Only from the 
anomalies in the first component, would deviations of 
the fission rate from its
transition state value be expected. 
The experimental
separation of the two contributions, however, is fraught
with difficulties which make the evidence ambiguous. 
It seems therefore desirable 
to search for transient time effects by directly measuring
the fission probability and its energy dependence against
the predictions of the transient state method for a large
number of systems and over a broad energy range. 

In the last few decades, several studies have investigated 
heavy ion and high energy light ion induced fission
\cite{Wag91}. 
These reactions involve a large and variable deposition 
of energy, mass and, most important, of angular momentum.
The latter, in particular, greatly affects the fission
process and makes comparisons with liquid drop
model calculations difficult \cite{Van73,Wag91}.
In contrast, the problem of excessive angular
momentum, mass and energy transfer and the associated 
uncertainties can be minimized 
by the use of light projectiles and relatively 
low bombarding energies, see e.g. Ref.~\cite{Rai67,Kho66,Iye91}.
Becchetti {\sl et al.} have, in particular, measured $^3$He induced
fission excitation functions of several nuclei with masses between 159 and
232 at bombarding energies ranging from 19.1 to 44.5 MeV
\cite{Bec83}. Their analysis with statistical 
fission theory indicates fission barriers
which, in contrast to heavy ion induced fission, 
differ only slightly from liquid drop model 
predictions.

In a recent letter, a new  
scaling of fission excitation functions based upon
the transition state prediction, collapses a large
number of fission excitation functions from
compound nuclei produced in $\alpha$-induced reactions 
\cite{Kho66} to
a single straight line, once the shell effects
are accounted for \cite{Mor95}. 
An investigation of fission delay times gave 
an upper limit of 3$\times$10$^{-20}$ seconds.

In this paper, we show the results of a recent
experiment investigating $^3$He induced fission of 
the compound nuclei $^{200}$Tl, $^{211}$Po, and $^{212}$At 
at excitation energies between 25 and 145 MeV. 
These fissioning systems bracket the closed shell region
around $^{208}$Pb, and due to the strong shell effects,
the analysis of these systems represents a sensitive 
test of the method introduced in Ref.~\cite{Mor95}.

The present paper is organized as follows:
In section~\ref{experiment}, we present an experimental setup 
which allows one to measure fission excitation 
functions for various nuclei efficiently, and we 
show the results of such a measurement.
In the subsequent section, we describe the analysis of the
fission data and our findings.
Finally, our summary can be found in section \ref{conclusion}.

\section{Experiment}
\label{experiment}
\begin{figure}[htb]
 \centerline{\psfig{file=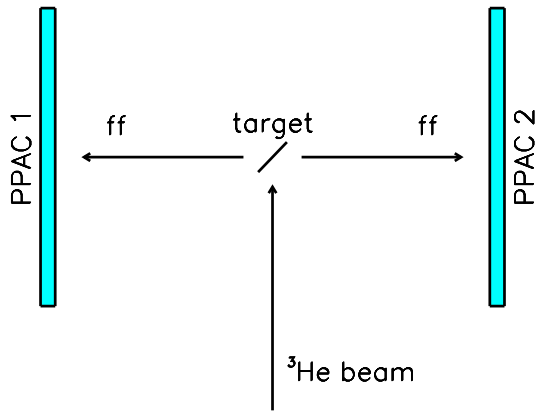,height=7cm}}
 \caption[]{Schematic view of the experimental setup. The beam
 enters from below, the fission fragments (ff) are detected in 
 coincidence  with the two PPACs mounted perpendicular to the beam.
 }
 \label{setup}
\end{figure}

Fission of three compound nuclei, 
$^{200}$Tl, $^{211}$Po, and $^{212}$At, formed in the reactions 
$^3$He + $^{197}$Au, $^{208}$Pb, and $^{209}$Bi was investigated. 
Fig.~\ref{setup} shows the schematic setup of the experiment.
The targets were mounted at 45 degrees with respect to the beam
axis and had thicknesses between 240 and 500~$\mu$g/cm$^2$.
The Lawrence Berkeley National
Laboratory's 88-Inch Cyclotron delivered $^{3}$He beams with 
19 different energies
between 21 MeV and 135~MeV. The number of energy points was 
increased to 26 
(see Table~\ref{xs}) 
by using a set of degraders made of aluminum foils
with thicknesses between 186 and 433~$\mu$m which were
determined by weighing.

In the past, these fission reactions have been studied using small
solid angle solid state counters or nuclear track detectors,
see e.g. Ref.~\cite{Rai67,Kho66,Iye91,Bec83}.
Therefore, beamtimes on the order of weeks were necessary
to measure complete excitation functions.
To cover a large solid angle and, therefore, to minimize
beam time, we performed an experiment using two 
large area parallel-plate-avalanche counters (PPACs)
with an active area of 200 x 240 mm$^2$ each. The detectors 
were mounted at 80$^{\circ}$ and 260$^{\circ}$ with respect to the beam axis,
allowing for the detection of both fission fragments in coincidence.
The PPACs were placed at a distances of 150~mm from the target
to the center of each detector.
As the beam energy increases the velocity of the compound nucleus 
in the  laboratory's frame increases, resulting
in a decreasing folding angle.
Since we require the detection of both fission fragments
in coincidence, and our detectors are mounted at a fixed
relative angle, the acceptance has a weak dependence on
the bombarding energy of the projectile:
For our detector setup,
we have determined a geometric angular coverage 
between 18 and 20\% for bombarding energies 
between 135 and 21~MeV, respectively.

The PPAC's detector volume is divided by a  cathode foil
made of 2~$\mu$m thick mylar foil which is set at a 
voltage of 450 - 550~V during operation. 
The readout of the cathode gives a position independent 
amplitude and time signal. On both sides of
the cathode, signal wireplanes are mounted at a distance 
of 3~mm, one with horizontal and the other one with vertical 
oriented wires. The wires have a thickness 
of 20~$\mu$m, the distance between the individual wires
is 1~mm. Five wires are combined to a group which
is read out by a delay line to reconstruct the
position of the particle.
An intrinsic resolution of 1.0~mm (FWHM) has been achieved in
both horizontal and  vertical position which allows for the
measurement of the folding angle precisely.
Each detector has an entrance window made
of mylar foil which separates the gas atmosphere in the detector
from the chamber vacuum. In the present experiment, the counters 
were operated by flowing isobutane gas at a constant 
pressure of 4~mbar. 

\begin{figure}[tb]
 \centerline{\psfig{file=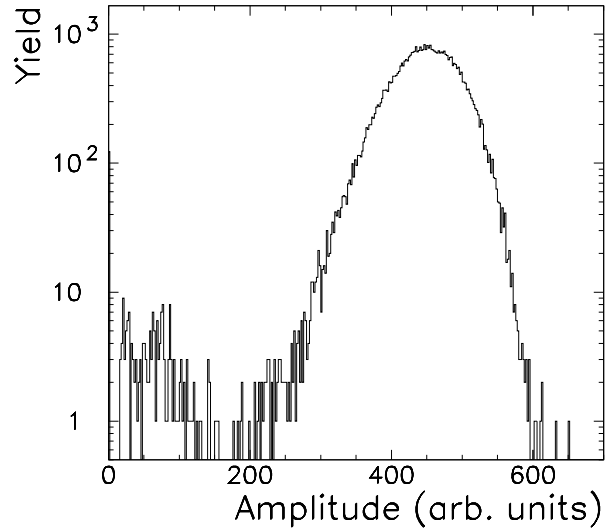,height=7.5cm}}
 \caption[]{Typical amplitude spectrum for coincidence events 
 as measured with the PPACs . 
 }
 \label{amp}
\end{figure}

In Fig.~\ref{amp}, we show a typical experimental amplitude spectrum 
for coincidence events. It shows that the fission peak is clearly 
visible and the background is negligible. 
Cross sections were determined for these fission events using
\begin{equation}
 \sigma_{f} = \frac{n_{f} A}{n_{beam} N_A m} \eta(\theta, \phi),
\end{equation}
where $n_f$ and $n_{beam}$ are the number of fission events and 
the number of beam particles, respectively. 
$A$ represents the mass number of the target, $N_A$ Avogadro's 
constant, and $m$ the thickness of the target. Due to the 
incomplete angular coverage, the quantity
$\eta(\theta,\phi)$ which accounts for the geometrical
acceptance and for the non-isotropic emission of the fission fragments 
has be be taken into account.
The anisotropic angular distribution 
$\frac{(d\sigma / d\Omega)_{\theta}}{(d\sigma / d\Omega)_{90^{\circ}}}$
of the fission fragments has been shown to be reasonably described 
by the function $\sin^{-1} \theta$ \cite{Van73}. 
We have used this dependence for the determination of our
acceptance.
The beam normalization was done using a Faraday cup. The 
systematic uncertainty  of this method can be estimated 
to $\pm$15\%. 

\begin{figure}[htb]
 \centerline{\psfig{file=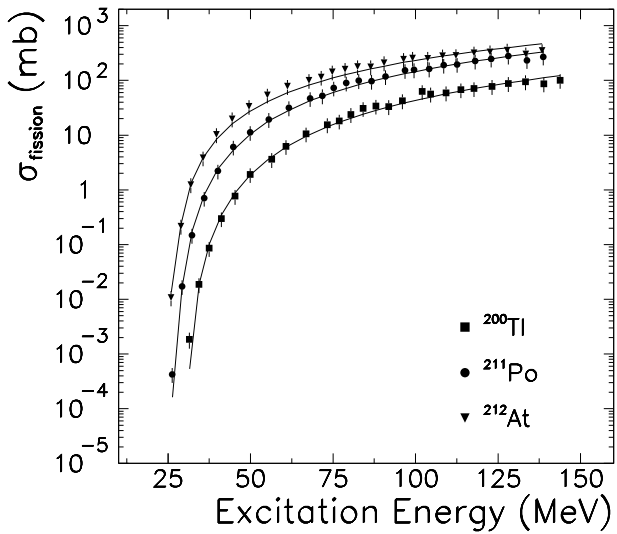,height=7.5cm}}
 \caption[]{Excitation function for fission of several compound
 nuclei formed in $^{3}$He induced reactions. The 
 different symbols correspond to the experimental data points.
 The solid line shows the results of a fit to the data using
 a level density parameter $a_{n} = A/8$.    
 The error bars denote the statistical and systematic errors
 combined in quadrature.
 }
 \label{exc_ftn}
\end{figure}

In Fig.~\ref{exc_ftn}, 
we show the experimental
fission cross sections for the three compound nuclei  
$^{200}$Tl, $^{211}$Po, and $^{212}$At as a function of 
excitation energy.
The error bars denote both the statistical and the systematic
errors. While the statistical errors dominates at the lowest 
energy points, the systematic uncertainties are the main
contribution at higher excitation energies.
The excitation energy was calculated assuming a full
momentum and mass transfer of the helium ions to the 
compound nucleus (CN). The binding energies of 
$^3$He, the target isotopes, and the compound nuclei
were taken from Ref.~\cite{Aud93}.

\begin{table}[htb]
 \caption[]{Experimental fission cross sections.}
 \begin{minipage}[htb]{3.4in}
 \renewcommand{\footnoterule}{\rule{1cm}{0cm} \vspace{-1.2cm}}
  \begin{tabular}{rr@{ $\pm$ }rr@{ $\pm$ }rr@{ $\pm$ }r}
    \noalign{\smallskip}\hline
   E($^3$He) &
   \multicolumn{2}{c}{ }&
   \multicolumn{2}{c}{$\sigma_{f}$ (mbarn)}  &
   \multicolumn{2}{c}{ } \\
   (MeV) &
   \multicolumn{2}{c}{$^{200}$Tl}  &
   \multicolumn{2}{c}{$^{211}$Po}  &
   \multicolumn{2}{c}{$^{212}$At} \\
    \noalign{\smallskip}\hline
   21.0	& 0.0019 & 0.0004 & 0.0004 & 0.0002 & 0.0108 & 0.0023\\
   24.0	& 0.0187 & 0.0038 & 0.0171 & 0.0036 & 0.2176 & 0.0441\\
   27.1	& 0.0859 & 0.0175 & 0.148  & 0.030 & 1.25 & 0.25\\
   30.8	& 0.298  & 0.060 & 0.707  & 0.141 & 3.8 & 0.8\\
   35.0	& 0.765  & 0.154 & 2.2    & 0.4 & 10.3 & 2.1\\
   39.7	& 1.9    & 0.4 & 6.1    & 1.2 & 19.9 & 4.0\\
   44.9	& 3.6    & 0.7 & 11.2   & 3.9 & 33.6 & 6.7\\
   50.6	& 6.2    & 1.3 & 19.4   & 6.3 & 54.2 & 10.9\\
   56.8	& 10.6   & 2.1 & 31.7   & 9.3 & 78.1 & 15.6\\
   63.4	& 15.5   & 3.1 & 46.6   & 10.5 & 100.6 & 20.2\\
   67.1$^a$	& 18.1   & 3.6 & 52.6   & 14.5 & 115.1 & 23.1\\
   70.6	& 23.9   & 4.8 & 72.7   & 17.9 & 143.8 & 28.8\\
   74.4$^a$	& 30.7   & 6.2 & 89.7   & 19.9 & 160.3 & 32.1\\
   78.3	& 34.1   & 6.8 & 99.7   & 19.4 & 178.6 & 35.8\\
   82.3$^a$	& 33.4   & 6.7 & 96.9   & 23.6 & 177.9 & 35.7\\
   86.5	& 42.6   & 8.5 & 117.8  & 30.6 & 211.3 & 42.3\\
   92.5	& 63.0   & 12.6 & 152.8  & 31.4 & 243.5 & 48.8\\
   95.2	& 56.6   & 11.3 & 157.0  & 32.6 & 255.8 & 51.3\\
   99.9$^a$	& 59.4   & 13.1 & 162.9  & 38.1 & 253.7 & 50.8\\
  104.4	& 67.4   & 14.9 & 190.3  & 39.3 & 282.4 & 56.6\\
  108.5$^a$	& 71.2   & 14.3 & 196.5  & 45.5 & 285.7 & 57.2\\
  114.1	& 78.1   & 15.6 & 227.3  & 49.2 & 318.3 & 63.7\\
  119.0$^a$	& 88.0   & 17.6 & 245.9  & 55.6 & 333.0 & 66.7\\
  124.3	& 94.9   & 19.0 & 277.8  & 46.3 & 358.9 & 71.9\\
  130.0$^a$	& 86.7   & 17.4 & 231.6  & 53.7 & 305.3 & 61.2\\
  135.0	& 100.3  & 20.1 & 268.6  & 54.3 & 351.5 & 70.4\\
    \noalign{\smallskip}\hline
   \end{tabular}
 \footnotetext[1]{The bombarding energy was achieved by using
 a degrader foil as described in the text.}
 \end{minipage}
 \label{xs}
\end{table}

\section{Analysis and results}
\label{method}
We will analyze our data according to 
a method introduced in Ref.~\cite{Mor95}
that allows us to investigate deviations 
from the transition state rates and 
enables us to extract effective fission barriers 
and values for the shell effects which are 
independent of those obtained from the ground state masses.
The transition state expression for the fission decay
width \cite{Wig38,Boh39}
\begin{equation}
 \Gamma_f \approx \frac{T_s}{2\pi} 
 \frac{\rho_s(E - B_f - E^s_r)}{\rho_n(E - E_r^{gs})} 
\end{equation}
allows one to write the fission cross section as follows: 
\begin{equation}
 \sigma_f = \sigma_0 \frac{\Gamma_f}{\Gamma_{total}}
 \approx \sigma_0 \frac{1}{\Gamma_{total}}
 \frac{T_s \rho_s (E - B_f - E^s_r)}{2\pi \rho_n (E - E_r^{gs})},
\end{equation}
where $\sigma_0$ is the compound nucleus formation cross section, 
$\Gamma_f$ is the decay width for fission 
and $T_s$ is the energy dependent temperature
at the saddle; $\rho_s$ and $\rho_n$ are the saddle and ground
state level densities, $B_f$ is the fission barrier,
and $E$ the excitation energy. Finally, $E^s_r$
and $E^{gs}_r$ represent the saddle and ground state rotational
energies.
This equation can be rewritten as
\begin{equation}
 \frac{\sigma_f}{\sigma_0} \Gamma_{total}
 \frac{2\pi \rho_n (E - E_r^{gs})}{T_s}
 = \rho_s (E - B_f - E_r^s).
\end{equation}
To further evaluate this expression, we use the form
$\rho(E) \propto \exp\big(2\sqrt{aE}\big)$ for the level density. 
This leads to:
\begin{equation}
 \ln\Big( \frac{\sigma_f}{\sigma_0} \Gamma_{total} 
 \frac{2\pi \rho_n (E - E_r^{gs})}{T_s} \Big) =
 2 \sqrt{a_f (E - B_f - E_r^s)}.
 \label{scal}
\end{equation}
If the transition state null hypothesis holds, plotting the left
hand side of the equation versus $\sqrt{E - B_f - E_r^s}$ 
should result in a straight line. This equation has already
been used in Ref.~\cite{Mor95a} to demonstrate the scaling of over 80 
excitation functions obtained by the study of the emission of 
complex fragments from compound nuclei like 
$^{75}$Br, $^{90,94}$Mo, and $^{110,112}$In.

Since the neutron width $\Gamma_n$
dominates the total decay width in our
mass and excitation energy regime, we can write:
\begin{equation}
 \Gamma_{total}  \approx \Gamma_n \approx K T_n^2 
 \frac{\rho_n(E - B_n - E_r^{gs})}{2\pi \rho_n(E - E_r^s)}
\end{equation}
where $B_n$ represents the binding energy of the last neutron,
$T_n$ is the temperature after neutron emission, and 
$K = \frac{2 m_n R^2 g'}{\hbar^2}$ with the spin degeneracy
$g'=2$.

The study of the fission process in the lead region
forces us to take strong shell
effects into account. For the fission excitation functions
discussed in this paper, the lowest excitation energies for
the residual nucleus after neutron emission are of the order of
15-20 MeV and therefore high enough to assume the asymptotic
form for the level density \cite{MorHui} which is given below:
\begin{eqnarray}
 \rho_n(E-B_n-E_r^{gs}) \propto 
\nonumber\\
 \exp \big(2 \sqrt{a_n(E-B_n-E_r^{gs}-\Delta_{shell})} \big)
\label{rho_n}
\end{eqnarray}
where $\Delta_{shell}$ is the ground state shell effect of the 
daughter nucleus ($Z,N-1$). 
For the level density at a few MeV above the saddle point, we can use
\begin{equation}
 \rho_s(E-B_f-E_r^{s}) \propto 
 \exp \big(2 \sqrt{a_f(E-B_f^*-E_r^{s})} \big)
 \label{rho_s}
\end{equation}
since the large saddle deformation implies small shell effects.
Deviations due to pairing, however, may be expected at very
low excitation energies. In Eq.~\ref{rho_s}, we introduced the
quantity $B_f^*$ which represents an effective fission barrier,
or, in other words, the unpaired saddle energy, i.e. 
$B_f^* = B_f + 1/2 g \Delta_0^2$ in the case of an even-even
nucleus and $B_f^* = B_f + 1/2 g \Delta_0^2 - \Delta_0$ for
nuclei with odd mass numbers. Here, $\Delta_0$ is the saddle
gap parameter and $g$ the density of doubly degenerate single
particle levels at the saddle.

\begin{table}[tb]
 \caption[]{Values of the effective fission barriers, $a_f/a_n$, 
  and shell effects. For comparison, we give the values 
  of the isotope $^{211}$Po obtained from the analysis of 
  $^4$He induced fission \protect\cite{Mor95} and
  the calculated shell effects $\Delta_{calc}$ taken
  from Ref.~\protect\cite{Mye94}. }
  \begin{tabular}{llr@{ $\pm$ }lr@{ $\pm$ }lr@{ $\pm$ }ll}
    \noalign{\smallskip}\hline
   CN &
   \multicolumn{1}{c}{Proj.} &
   \multicolumn{2}{c}{$B_f^*$ (MeV)} &
   \multicolumn{2}{c}{$a_f/a_n$} &
   \multicolumn{2}{c}{$\Delta_{shell}$ (MeV)}&
   \multicolumn{1}{c}{$\Delta_{calc}$ (MeV)} \\
    \noalign{\smallskip}\hline
    $^{212}$At & $^3$He & 19.5 & 1.0 & 1.008 & 0.020 & 10.7 & 1.5 & 9.6\\
    $^{211}$Po & $^3$He & 23.0 & 1.0 & 1.009 & 0.030 & 13.7 & 1.5 & 10.8\\
    $^{211}$Po & $^4$He & 23.1 & 1.5 & 1.028 & 0.050 & 13.4 & 1.5 & 10.8\\
    $^{200}$Tl & $^3$He & 25.1 & 1.0 & 0.995 & 0.046 & 12.1 & 1.5 & 6.6\\
    \noalign{\smallskip}\hline
   \end{tabular}
 \label{res}
\end{table}

Finally, the use of Eq.~\ref{rho_n} and Eq.~\ref{rho_s} 
for the level densities allows us to study the scaling of the
fission probability as introduced in Eq.~\ref{scal}:
\begin{eqnarray}
 \frac{1}{2 \sqrt{a_n}} \ln \Big(\frac{\sigma_f}{\sigma_0}
 \Gamma_{total} \frac{2\pi\rho_n(E-E_r^{gs})}{T_s}\Big) =
\nonumber\\
 \frac{\ln R_f}{2\sqrt{a_n}}
 = \sqrt{\frac{a_f}{a_n}(E - B_f^* - E_r^s)}.
\label{rf_eq}
\end{eqnarray}
The values for $B_f^*$, $\Delta_{shell}$, and $a_f/a_n$ using
$a_n = A/8$ can be obtained by a three parameter fit of
the experimental fission excitation functions;
the best results of the fits are shown in
Fig.~\ref{exc_ftn} and listed in Table~\ref{res}.
For this procedure, the formation cross sections $\sigma_0$ 
and the corresponding values for the maximum angular momentum
$l_{max}$ were taken from an optical model calculation \cite{Pto}. 
A simple parametrization, $\sigma_0 = \sigma_{geom}(1 - V/E_{cm})$,
where $\sigma_{geom}$ is the geometrical cross section, $V$ the
Coulomb barrier, and $E_{cm}$ the energy in the center of mass,
was used to interpolate the results of the 
optical model calculations.
Here, we used the expressions $V = (Z_1 Z_2 e^2)/R$ for the
Coulomb barrier, $R = r_0 (A_1^{1/3} + A_1^{1/3} + \delta)$,
and $\sigma_{geom} = 2\pi R^2$ for the geometrical 
cross section. The parameters $r_0$ and $\delta$ were chosen
so that the resulting cross sections are in agreement
with the optical model calculations.
The overall uncertainty of the calculated formation cross
sections can be estimated to 5\%.
Finally, we computed the rotational energy at the saddle
assuming a configuration of two nearly touching spheres
separated by 2~fm.
 
In a previous letter, it has been shown that
the employed method allows one to
extract values for the shell effect directly from the
data in contrast to the standard procedure where 
shell effects are determined by the difference of 
the ground state mass and the corresponding liquid drop
value \cite{Mor95}. 
Furthermore, it has been pointed out
that the determination of the 
shell effects is completely local since it only depends 
on the properties of the considered nucleus.

\begin{figure}[htb]
 \centerline{\psfig{file=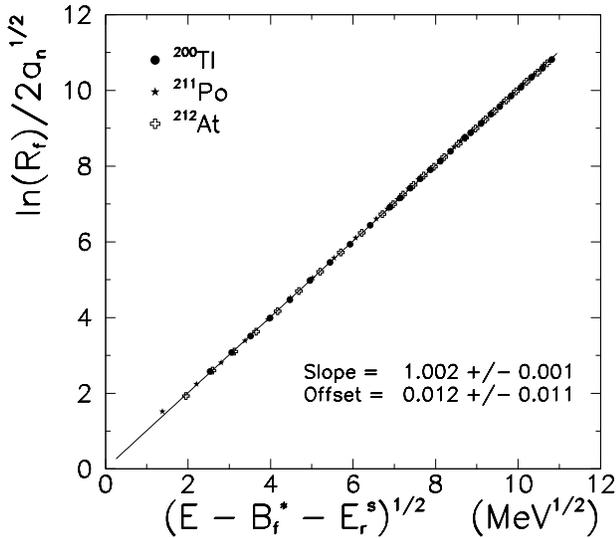,height=7.5cm}}
 \caption[]{The quantity $\frac{\ln R_f}{2 \sqrt{a_n}}$ vs 
 the square root of the intrinsic excitation energy over the 
 saddle for fission of several compound nuclei as described
 in the text. The straight line represents a fit to the entire
 data set.
 }
 \label{rf}
\end{figure}

In Fig.~\ref{rf}, we now plot the left hand side of Eq.~\ref{rf_eq}
versus the square root of the effective excitation energy
above the barrier, $\sqrt{E-B_f^*-E_r^s}$, including the results
of the fits described above. We should note that we {\it do not}
make use of the fitted value of $a_f/a_n$.
A remarkable straight line can
be observed for the three investigated compound nuclei.
This scaling extends over six orders of magnitude in the
fission probability, although the shell effects are 
very strong in this regime. 
Furthermore, a linear fit to the data results in a straight line 
that goes through the origin and has 
a slope which represents the ratio $a_f/a_n$,  
consistent with unity. 
The observed scaling and the lack of deviations
over the entire range of excitation energy  
indicates that the transition state null hypothesis 
and the above discussed equations for the level
density hold very well. 
The result of this work is in complete agreement
with the findings of a similar analysis investigating
14 $\alpha$-induced fission excitation functions 
\cite{Mor95,Rub96a}.

The presentation of the experimental data in Fig.~\ref{rf}
and Eq.~\ref{scal} implies the dominance of first chance fission. 
Calculations verify that first chance fission dominates completely
at the lower energies.
Even for the highest energy range,
first chance fission still accounts for a large part
of the cross sections. However, some uncertainties 
with the nuclear parameters, such as the barriers,
shell effects occur for the higher chance 
fissioning nuclei. 
It certainly remains an interesting question to 
experimentally investigate first chance fission probabilities
with an appropriate accuracy and to apply the results to
the method introduced in Ref.~\cite{Mor95}.

\begin{figure}[tb]
 \centerline{\psfig{file=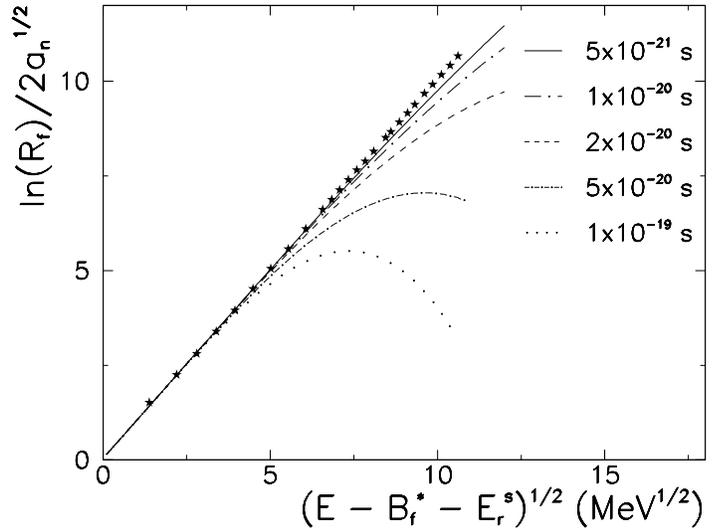,height=7.5cm}}
 \caption[]{Same as Fig.~\protect\ref{rf} for $^{211}$Po (stars). 
 The lines
 represent calculations assuming that no fission occurs
 during a given transient time which is indicated in the
 figure. For further details see text.
 }
 \label{time}
\end{figure}

The excitation energy range covered by our experiment
corresponds to life times of the compound nuclei between 
10$^{-18}$ and 10$^{-22}$ seconds, and should therefore
be sensitive to delay times in the fission process.
To investigate this effect, we assume a step
function for the transient time effects. In this assumption, 
the fission width can be written as follows:
\begin{equation}
 \Gamma_f = \Gamma_f^{\infty} \int^{\infty}_{0} \lambda(t)
 \exp(\frac{-t}{\tau_{CN}}) d(\frac{t}{\tau_{CN}}) = 
 \Gamma_f^{\infty} \exp(\frac{-\tau_D}{\tau_{CN}})
\end{equation}
where the quantity $\lambda(t)$ jumps from 0 at times smaller
than the transient time $\tau_D$ to 1 for times larger than
$\tau_D$. Furthermore, $\Gamma_f^{\infty}$ denotes the
transition state fission decay width and $\tau_{CN}$ 
represents the life time of the compound nucleus. 
This expression for the fission decay width has been
used in the formalism described above; the parameters
$B_f^*$, $\Delta_{shell}$, and $a_f/a_n$ have been
taken from Table~\ref{res}.
In Fig.~\ref{time}, we show the results of these 
calculations for the compound nucleus $^{211}$Po; the 
different lines indicate 
different assumed values of the 
transient time between 1$\times$10$^{-19}$ and 5$\times$10$^{-21}$
seconds. 
The calculated values show an obvious deviation
from the experimental data for assumed
transient times larger than 10$^{-20}$ seconds. 
Since the experimental fission rates are well described
by the transition state rates, 
it seems likely that any excess prescission emission 
occurs during the descent  from saddle to scission.
If this is the case, then the present fission results
are not in contradiction with recent measurements of prescission 
neutrons and $\gamma$ rays \cite{Hil92,Pau94,Tho93}.

\section{Summary}
\label{conclusion}
Experimentally, we have investigated $^3$He induced 
fission excitation functions of three different
compound nuclei, $^{200}$Tl, $^{211}$Po, and $^{212}$At
between 25 and 140 MeV excitation energy.

The data have been analyzed and discussed according to
a method which allows one to check 
the validity of the transition state null hypothesis over a large
range of excitation energy and a regime of compound nuclei masses
which is characterized by strong shell effects.  
Once these shell effects are accounted for, no deviation from
the transition state rate is observed. 
Furthermore, the shell effects can be determined directly 
from the experimental data by using the above described procedure. 
Finally, plotting the reduced fission rate $R_f$ allows one 
to look for evidence of fission delay times as they have been discussed 
in a series of papers. Our results, however, indicate that 
the proposed transient 
times -- if they exist -- are shorter than 10$^{-20}$ seconds.

\bigskip

{\noindent\bf Acknowledgement}\\
The authors would like to thank the staff of the 88-Inch Cyclotron
for the excellent support, i.e. the fast changes of beam energies. 
Discussions with M.-E. Brandon are gratefully acknowledged.
This work was supported by the Director, Office of Energy Research,
Office of High Energy and Nuclear Physics, Nuclear Physics Division
of the US Department of Energy, under contract DE-AC03-76SF00098.


\begin{thebibliography}{References}{}

\bibitem{Fis}
O. Hahn and F. Strassmann, Naturwiss. {\bf 26}, 756 (1938),
O. Hahn and F. Strassmann, Naturwiss. {\bf 27}, 11 (1939),
L. Meitner and O.R. Frisch, Nature {\bf 3615}, 239 (1939).

\bibitem{Sch94}
K.H. Schmidt, A. Heinz, H.-G. Clerc, B. Blank, T. Brohm, 
S. Czajkowski, C. Donzaud, H. Geissel, E. Hanelt, H. Irnich, 
M.C. Itkis, M. de Jong, A. Junghans, A. Magel, G. M\"unzenberg,
F. Nickel, M. Pf\"utzner, A. Piechaczek, C. R\"ohl, C. Scheidenberger, 
W. Schwab, S. Steinh\"auser, K. S\"ummerer, W. Trinder, B. Voss, 
S.V. Zhdanov,
Phys. Lett. B{\bf\,325}, 313 (1994).

\bibitem{Ber94}
M. Bernas, S. Czajkowski, P. Armbruster, H. Geissel, Ph. Dassagne, 
C. Donzaud, H.-R. Faust, E. Hanelt, A. Heinz, M. Hesse, C. Kozhuharov, 
Ch. Miehe, G. M\"unzenberg, M. Pf\"utzner, C. R\"ohl, K.-H. Schmidt, 
W. Schwab, C. St\'{e}phan, K. S\"ummerer, L. Tassan-Got, B. Voss, 
Phys. Lett. B{\bf\,331}, 19 (1994).
                                           
\bibitem{Pol94}
S. Polikanov, W. Br\"uchle, H. Folger, E. J\"ager, T. Krogulski,
M. Sch\"adel, E. Schimpf, G. Wirth, T. Aumann, J.V. Kratz,
and N. Stiel, N. Trautmann,
Z. Phys. A {\bf\, 350}, 221 (1994).

\bibitem{Cle95}
H.-G. Clerc, M. de Jong, T. Brohm, M. Dornik, A. Grewe,
E. Hanelt, A. Heinz, A. Junghans, C. R\"ohl, S. Steinh\"auser,
B. Voss, C. Ziegler, K.-H. Schmidt, S. Czajkowski, H. Geissel,
H. Irnich, A. Magel, G. M\"unzenberg, F. Nickel, A. Piechaczek,
C. Scheidenberger, W. Schwab, K. S\"ummerer, W. Trinder,
M. Pf\"utzner, B. Blank, A. Ignatyuk, and G. Kudyaev,
Nucl. Phys. A{\bf\, 590}, 785 (1995).

\bibitem{Ign95}
A.V. Ignatyuk, G.A. Kudyaev, A. Junghans, M. de Jong,
H.-G. Clerc, and K.-H. Schmidt,
Nucl. Phys. A{\bf 593}, 519 (1995)
  
\bibitem{Rub95}
Th. Rubehn, W.F.J. M\"uller, R. Bassini, M. Begemann-Blaich,
Th. Blaich, A. Ferrero, C. Gro\ss, G. Imm\'{e}, I. Iori,
G.J. Kunde,  W.D. Kunze, V. Lindenstruth, U. Lynen,
T. M\"ohlenkamp, L.G. Moretto, B. Ocker,  J. Pochodzalla,
G. Raciti, H. Sann,  A. Sch\"uttauf, W. Seidel, V. Serfling,
W. Trautmann, A. Trzcinski,
G. Verde, A. W\"orner, E. Zude, and B. Zwieglinski,
Z. Phys. A{\bf\, 353}, 197 (1995); 
Phys. Rev. C{\bf\,53}, 993 (1996);
Phys. Rev. C (in print).

\bibitem{Arm95}
P. Armbruster, M. Bernas,
T. Aumann, S. Czajkowski,
H. Geissel, Ph. Dessagne, C. Donzaud,
E. Hanelt, A. Heinz, M. Hesse,
C. Kozhuharov, Ch. Miehe, G. M\"unzenberg,
M. Pf\"utzner, K.-H. Schmidt, W. Schwab,
C. Stephan, K. S\"ummerer, L. Tassan-Got, and B. Voss,
Z. Phys. A (submitted).

\bibitem{Mor95} 
L.G. Moretto, K.X. Jing, R. Gatti, R.P. Schmitt, and G.J. Wozniak,
Phys. Rev. Lett.{\bf~75}, 4186 (1995).

\bibitem{Hil92} 
D. Hilscher and H. Rossner, Ann. Phys. Fr.{\bf~17}, 471 (1992).

\bibitem{Pau94}
P. Paul and M. Thoennessen, Ann. Rev. Nucl. Part. Sc.{\bf~44},
65 (1994).

\bibitem{Rai67}
G. M. Raisbeck and J.W. Cobble,
Phys. Rev. {\bf~153}, 1270 (1967).

\bibitem{Mor72}
L.G. Moretto, S.G. Thompson, J. Routti, and R.C. Gatti,
Phys. Lett.{\bf~38B}, 471 (1972).

\bibitem{Kho66}
A. Khodai-Joopari, 
Ph.D. thesis, University of California at Berkeley, 1966.

\bibitem{Van73}
R. Vandenbosch, J.R. Huizenga, 
{\sl Nuclear Fission} 
(Academic Press, New York, 1973)
and references therein.

\bibitem{Wag91}
C. Wagemans, {\sl The Nuclear Fission Process} 
(CRC Press, Boca Raton - Ann Arbor - Boston - London, 1991)
and references therein.

\bibitem{Wig38}
E. Wigner, Trans. Faraday Soc.{\bf~34}, 29 (1938).

\bibitem{Boh39}
N. Bohr and J.A. Wheeler, Phys. Rev.{\bf~56}, 426 (1939).

\bibitem{Tho93}
M. Thoennessen and G.F. Bertsch, Phys. Rev. Lett.{\bf~71}, 4303 (1993).

\bibitem{Gra83a}P. Grang\'{e} and H.A. Weidenm\"uller,
Phys. Lett.{\bf~B96}, 26 (1980).

\bibitem{Gra83b}P. Grang\'{e}, L. Jun-Qing, and H.A. Weidenm\"uller,
Phys. Rev. C{\bf~27}, 2063 (1983).

\bibitem{Wei84}H.A. Weidenm\"uller and Z. Jing-Shang,
Phys. Rev. C{\bf~29}, 879 (1984).

\bibitem{Gra86}P. Grang\'{e}, S. Hassani, H.A. Weidenm\"uller,
A. Gavron, J.R. Nix, and A.J. Sierk, 
Phys. Rev. C{\bf~34}, 209 (1986).

\bibitem{Lu86}L. Zhongdao, Z. Jingshang, F. Renfa, and
Z. Yizhong,
Z. Phys. A{\bf~323}, 477 (1986).

\bibitem{Lu90}Z.-D. Lu, B. Chen, 
J.-S. Zhang, Y.-Z. Zhuo,
and H.-Y. Han, 
Phys. Rev. C{\bf~42}, 707 (1990).
   
\bibitem{Cha92}
D. Cha and G.F. Bertsch, 
Phys. Rev. C{\bf~46}, 306 (1992).

\bibitem{Fro93}
P. Fr\"obrich, I.I. Gontchar, and N.D. Mavlitov, Nucl. Phys. A{\bf~556},
281 (1993).

\bibitem{Siw95}
K. Siwek-Wilczy\'{n}ska, J. Wilczy\'{n}ski, R.H. Siemssen,
and H.W. Wilschut, 
Phys. Rev. C{\bf~51}, 2054 (1995).
\bibitem{Iye91}
R.H. Iyer, A.K. Pandey, P.C. Kalsi, and C. Sharma,
Phys. Rev. C{\bf~44}, 2644 (1991).

\bibitem{Bec83}
F.D. Becchetti, K.H. Hicks, C.A. Fields, R.J. Peterson,
R.S. Raymond, R.A. Ristinen, J.L. Ullmann, and C.S. Zaidins, 
Phys. Rev. C{\bf~28}, 1217 (1983).

\bibitem{Aud93}
G. Audi and A.H. Wapstra, Nucl. Phys. A{\bf~565}, 1 (1993).

\bibitem{Mor95a} 
L.G. Moretto, K.X. Jing, and G.J. Wozniak, 
Phys. Rev. Lett.{\bf~74}, 3557 (1995).

\bibitem{MorHui}
J.R. Huizenga and L.G. Moretto, 
Ann. Rev. Nucl. Sci. {\bf 22}, 427 (1972).

\bibitem{Mye94}
W.D. Myers and W.J. Swiatecki, 
{\it Table of nuclear masses according to the 1994
Thomas - Fermi model},
(Lawrence Berkeley National Laboratory, LBL-36803, 1994).

\bibitem{Pto}
Optical model code {\sc ptolemy},
M.H. Macfarlane and S.C. Pieper, 
(Argonne National Laboratory, ANL-76-11, 1978).

\bibitem{Rub96a}
Th. Rubehn, K.X. Jing, L.G. Moretto, L. Phair, K. Tso, and
G.J. Wozniak, in {\it Proceedings of the 12th Winter Workshop on
Nuclear Dynamics}, Snowbird, Utah, 1996, edited by W. Bauer and
G.D. Westfall 
[Plenum Press, 1996] (in press)


\end{thebibliography}
\end{document}